# From Random Motion of Hamiltonian Systems to Boltzmann's *H* Theorem and Second Law of Thermodynamics: a Pathway by Path Probability

Qiuping A. Wang [1,2]* and Aziz El Kaabouchiu [1]

1. Laboratoire de Physique Statistique et Systems Complexes, ISMANS, 44 Ave. F.A., Bartholdi, 72000 Le Mans, France; E-Mail : aek@ismans.fr
2. IMMM, Université du Maine, Ave. O. Messiaen, Le Mans 72085, France

* Author to whom correspondence should be addressed; E-mail: awang@ismans.fr; Tel.: +33-243-214-026; Fax: +33-243-214-039.

**Abstract:** A numerical experiment of ideal stochastic motion of a particle subject to conservative forces and Gaussian noise reveals that the path probability depends exponentially on action. This distribution implies a fundamental principle generalizing the least action principle of the Hamiltonian/Lagrangian mechanics and yields an extended formalism of mechanics for random dynamics. Within this theory, Liouville's theorem of conservation of phase density distribution must be modified to allow time evolution of phase density and consequently the Boltzmann *H* theorem. We argue that the gap between the regular Newtonian dynamics and the random dynamics was not considered in the criticisms of the *H* theorem.

## 1. Introduction

It is well known that for regular motion obeying Newtonian mechanics, the path between two given points in configuration space as well as in phase space when the time period is specified is unique [1]. However, it is not the case for random dynamics with a well-known example: Brownian motion. One of the remarkable characteristics of this motion is the non uniqueness of paths between two given points for given time period, which is illustrated in Figure 1 (left).

**F**igure 1: An example of the results of numerical simulation of random motion with $10^9$ particles subject to a harmonic force f= $-kx$ and Gaussian noise, where *x* is the position of the particle with respect to the initial point *a*. The left panel shows about 100 paths between two given points. The duration of motion is about three halves of a period of the harmonic oscillation. These paths are created through the Gaussian noise around the most probable path which is just the harmonic curve between the points *a* and *b*. The magnitude of the noise is controlled in order that the paths are sufficiently far from each other to give sufficiently different energies and actions but not too far to make the paths sufficiently smooth. The right panel shows the correlation of the probability distribution of paths $p(A_k)$ with action (o), compared to the uncorrelation with Hamiltonian action (*). The most probable paths are the paths of least action which is just the ballistic path of the harmonic oscillation. The straight line in the right panel is just guide for eye.

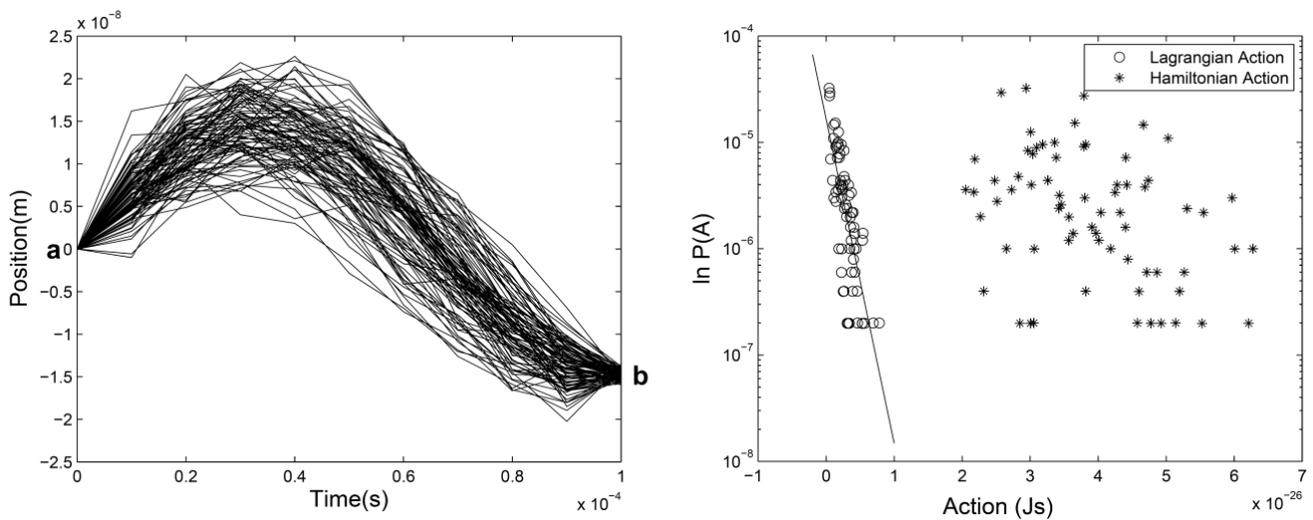

This phenomenon naturally raises the following questions. What is the probability for a particle, moving between two points, to take a given path or to follow a thin bundle of paths among many others? What are the variables of this probability? How to quantify the dynamical uncertainty in the path probability? Although the path probability of random motion has been noticed since longtime ago [2,3], explicit study of the expression of the path probability and its random variables is relatively new. There are different viewpoints on the question [4–7] which do not always agree with each other. In a recent work [8], the path probability of an ideal random motion was studied by numerical simulation with particles subject to conservative force and Gaussian noise. This model is ideal since the random motion statistically conserves its mechanical energy in such a way that the Hamiltonian can make sense, such as in a thermodynamically isolated system with internal fluctuation. In practice, any mechanical system with weak dissipation (or friction) can be approximately described by this model within the period in which the energy dissipated is much smaller than the energy of the motion. Figure 1 gives an example of the results. The exponential dependence of the path probability on the action is obvious from this result. The action is denoted by $A_k(a,b)$ for a given path *k* between two fixed points *a* and *b* with the usual definition $A_k(a,b) = \int_a^b (K-V)_k dt$ where *K* is the kinetic energy and *V* the potential energy along the path. In that work, we called it Lagrangian action to distinguish it from the integral of Hamiltonian $A_H = \int_a^b (K+V)_k dt$ along the same path (called Hamiltonian action in that work) used to compare the action and energy correlation of the path probability. The result in the right panel of Figure 1 can be expressed by:

$$p_k(a,b) = \frac{1}{Z_{ab}} e^{-\gamma A_k(a,b)} \tag{1}$$

where $\gamma$ is the slope of the straight line in the right panel and $Z_{ab}$ is the partition function given by $Z_{ab} = \sum_k e^{-\gamma A_k(a,b)}$ summed over all the possible paths between *a* and *b*. In general, this summation can be carried out by path integral [9].

In this paper, we present some theoretical consequences of this numerical result. The main point is that this result implies a modification of Liouville's theorem of conservation of phase volume due to the modification of Hamiltonian equations, which results in other important consequences relative to the entropy of thermodynamics.

## 2. Stochastic Least Action Principle

Let's first see the probabilistic uncertainty, or randomness, in the choice of paths by the motion. This uncertainty is represented by the distribution Equation (1) and can be measured by the Shannon information formula $S_{ab} = -\sum_k p_k(a,b) \ln p_k(a,b)$. Let's call it path entropy. It is easy to calculate:

$$S_{ab} = \ln Z_{ab} + \gamma \overline{A}_{ab} \tag{2}$$

On the other hand, the distribution of Equation (1) maximizes the path entropy. In other words, the vanishing variation $\delta(S_{ab} - \gamma \overline{A}_{ab}) = 0$ yield Equation (1). Considering that $\delta \ln Z_{ab} = \sum_k p_k(a,b) \delta A_k = \overline{\delta A}$ and Equation (2), the exponential path probability Equation (1) implies:

$$\overline{\delta A} = 0 \tag{3}$$

where $\delta A$ is a variation of the Lagrangian action and the average is over all possible paths between *a* and *b*. This formula has been called stochastic least action principle (SAP) as proposed in our previous work [10–13]. Here we have shown that it is in accordance with the distribution of Equation (1), a result of the numerical experiment of random motion of Hamiltonian (non-dissipative) systems.

## 3. Hamiltonian Mechanics Revisited

It has been shown that the exponential probability distribution of action satisfied the Fokker-Planck equation for normal diffusion in the same way as the Feymann factor of quantum propagator $P_k \propto e^{-iA_k/\hbar}$ satisfies the Schrödinger equation [9] Here we will show a generalized formalism of Hamiltonian mechanics whose equations will be used in the discussion of the Liouville theorem and Boltzmann *H* theorem.

*3.1. Euler-Lagrange Equations*

A meaning of the SAP Equation (3) is that for any particular path *k*, there is no necessarily $\delta A_k = 0$. In general we have:

$$\delta A_k = \int_a^b \left[ \frac{\partial}{\partial t}\left(\frac{\partial L_k}{\partial \dot{x}}\right) - \frac{\partial L_k}{\partial x} \right] \delta x \, dt \geq \text{ (or } \leq\text{) } 0 \tag{4}$$

where $\delta x$ is an arbitrary variation of *x* which is zero at *a* and *b*. For $\delta A_k \geq 0$ (or $\leq 0$), we get:

$$\frac{\partial}{\partial t}\left(\frac{\partial L_k}{\partial \dot{x}}\right) - \frac{\partial L_k}{\partial x} \geq \text{ (or } \leq\text{) } 0 \tag{5}$$

which can be proved by contradiction as follows. Suppose $\int_a^b f(t)\delta x \, dt \geq 0$ and $f(t) \leq c \leq 0$ during a small period of time $\Delta t$ somewhere between *a* and *b*. Since $\delta x$ is arbitrary, let it be zero outside $\Delta t$ and a positive constant within $\Delta t$. We clearly have $\int_a^b f(t)\delta x \, dt \leq c\delta x \leq 0$, which contradicts our starting assumption. This proves Equation (5).

The Legendre transformation $H_k = P_k \dot{x} - L_k$ along a path *k* implies the momentum given by $P_k = \frac{\partial L_k}{\partial \dot{x}}$ which can be put into Equation (5) to have:

$$\dot{P}_k \geq \text{ (or } \leq\text{) } \frac{\partial L_k}{\partial x} \tag{6}$$

for $\delta A_k \geq (or \leq) 0$.

However, from the path average of Equation (4) and the SAP $\overline{\delta A} = 0$, we straightforwardly write:

$$\overline{\delta A} = \sum_k p_k \int_a^b \left[ \frac{\partial}{\partial t}\left(\frac{\partial L_k}{\partial \dot{x}}\right) - \frac{\partial L_k}{\partial x} \right] \delta x \, dt = \int_a^b \left[ \overline{\frac{\partial}{\partial t}\left(\frac{\partial L_k}{\partial \dot{x}}\right)} - \overline{\frac{\partial L_k}{\partial x}} \right] \delta x \, dt = 0 \tag{7}$$

which implies:

$$\overline{\frac{\partial}{\partial t}\left(\frac{\partial L_k}{\partial \dot{x}}\right)} - \overline{\frac{\partial L_k}{\partial x}} = 0 \tag{8}$$

This is the Euler-Lagrange equation of the random dynamics. We have equivalently:

$$\overline{\dot{P}} = \overline{\frac{\partial L}{\partial x}} \tag{9}$$

where $\overline{\dot{P}} = \sum_k p_k \dot{P}_k$ and $L = \sum_k p_k L_k$.

*3.2. Hamiltonian Equations*

From Legendre transformation, we can have $\frac{\partial L_k}{\partial x} = -\frac{\partial H_k}{\partial x}$ and the following Hamiltonian equations:

$$\dot{x}_k = \frac{\partial H_k}{\partial P_k} \text{ and } \dot{P}_k \geq (or \leq) - \frac{\partial H_k}{\partial x} \qquad (10)$$

For a path $k$ along which $\delta A_k \geq (or \leq) 0$.

Naturally, Equation (8) means:

$$\overline{\dot{P}} = \overline{\frac{\partial H}{\partial x}} \qquad (11)$$

with the average Hamiltonian $H = \sum_k p_k H_k$.

## 4. Liouville's Theorem

Liouville's theorem is often involved in the discussions relative to thermodynamic entropy in statistical mechanics. We give an outline below followed by an analysis of the theorem within the present formulation of (probabilistic) Hamiltonian mechanics.

We look at the time change of phase point density $\rho(x, P, t)$ in a, say, 2-dimensional phase space $\Gamma$ when the system of interest moves on the path of least action [3] $\rho(x, P, t)$ can also be considered as the density of systems of an ensemble of a large number of systems moving in phase space. The time evolution neither creates nor destroys state points or systems, hence the law of state conservation in the phase space is:

$$\frac{\partial \rho}{\partial t} + \frac{\partial (\dot{x}\rho)}{\partial x} + \frac{\partial (\dot{P}\rho)}{\partial P} = 0 \qquad (12)$$

which means:

$$\frac{d\rho}{dt} = \frac{\partial \rho}{\partial t} + \frac{\partial \rho}{\partial x}\dot{x} + \frac{\partial \rho}{\partial P}\dot{P} = -\left(\frac{\partial \dot{x}}{\partial x} + \frac{\partial \dot{P}}{\partial P}\right)\rho \qquad (13)$$

For the least action path satisfying Hamiltonian equations [1], the right hand side of the above equation is zero, leading to the Liouville's theorem:

$$\frac{d\rho}{dt} = 0 \qquad (14)$$

*i.e.*, the state density in phase space is a constant of motion. The phase volume $\Omega$ available to the system can be calculated by $\Omega = \int_\Gamma \frac{1}{\rho} dn$ where $dn$ is the number of phase point in an elementary volume $d\Gamma$ at some point in phase space. The time evolution of the phase volume $\Omega$ accessible to the system is then given by:

$$\frac{d\Omega}{dt} = \frac{d}{dt}\int_\Gamma \frac{1}{\rho} dn = -\int_\Gamma \frac{1}{\rho^2}\frac{d\rho}{dt} dn = 0 \qquad (15)$$

meaning that this phase volume is a constant of motion.

The second law of thermodynamics states that the entropy of an isolated system increases or remains constant in time. But Liouville's theorem implies that if the motion of the system obeys the fundamental laws of mechanics, the Boltzmann entropy defined by $S = \ln \Omega$ must be constant in time. On the other hand, the probability distribution of states $p(x,P)$ in phase space is proportional to $\rho(x,P)$, meaning that an entropy $S(p)$, as a functional of $p(x,P)$, must be constant in time, which is in contradiction with the second law.

What is then the Liouville's theorem in the context of random dynamics? From Equation (13), we have $\left.\dfrac{d\rho}{dt}\right|_k = -\left(\dfrac{\partial \dot{x}_k}{\partial x} + \dfrac{\partial \dot{P}_k}{\partial P_k}\right)\rho$ along a path $k$. Let us write the second equation of the Equations (10) as $\dot{P}_k = -\dfrac{\partial H_k}{\partial x} + R_k$ where $R_k \geq (or \leq) 0$ for $\delta A_k \geq (or \leq) 0$ is the random force causing the deviation from Newtonian laws. The average Newtonian law $\overline{\dot{P}} = \dfrac{\overline{\partial H}}{\partial x}$ yields $\sum_k p_k R_k = 0$ for any moment of the process. We then have:

$$\left.\frac{d\rho}{dt}\right|_k = -\frac{\partial R_k}{\partial P_k}\rho \qquad (16)$$

and:

$$\frac{d\rho}{dt} = \sum_k p_k \left.\frac{d\rho}{dt}\right|_k = -\overline{\frac{\partial R_k}{\partial P_k}}\rho \qquad (17)$$

where $\overline{\dfrac{\partial R_k}{\partial P_k}} = \sum_k p_k \dfrac{\partial R_k}{\partial P_k}$ is an average over all the possible paths. The solution of this equation is:

$$\rho(t) = \rho(t_0)\exp[\zeta(t,t_0)] \qquad (18)$$

with the function $\zeta(t,t_0) = -\int_{t_0}^{t} \overline{\dfrac{\partial R_k}{\partial P_k}} dt$.

Now let us see an application to the case of exponential path probability. The relationship $\sum_k p_k R_k = 0$ implies $\overline{\dfrac{\partial R_k}{\partial P_k}} = -\sum_k R_k \dfrac{\partial p_k}{\partial P_k}$. By using the exponential distribution of action, we have $\dfrac{\partial p_k}{\partial P_k} = -\gamma p_k \dfrac{\partial A_k}{\partial P_k}$ where $\dfrac{\partial A_k}{\partial P_k} = \int_{t_0}^{t}\dfrac{\partial L_k}{\partial P_k}d\tau = \int_{t_0}^{t}\dot{x}_k d\tau = x_k(t)$ (suppose $x(t_0) = 0$) where $x_k(t)$ is the displacement of the motion along the path $k$ from $t_0$ to $t$. Hence $\overline{\dfrac{\partial R_k}{\partial P_k}} = \gamma \sum_k p_k \int_{t_0}^{t} R_k \dot{x}_k d\tau = \gamma \overline{W(t)}$ where $\overline{W(t)} = \sum_k p_k W_k(t)$ is the average of the work $W_k(t) = \int_{t_0}^{t} R_k \dot{x}_k d\tau$ performed by the random forces $R_k$ (independent from time) over the displacement $x_k(t)$. Finally, $\zeta(t,t_0) = -\gamma \int_{t_0}^{t}\overline{W(\tau)}d\tau = -\gamma W(t,t_0)$ here $W(t,t_0) = \int_{t_0}^{t}\overline{W(\tau)}d\tau$ is the cumulate average work of the random force performed from $t_0$ and $t$. We have:

$$\rho(t) = \rho(t_0)\exp[-\gamma W(t,t_0)] \qquad (19)$$

meaning that the state density decreases (increases) and the phase volume increases (decreases) whenever the cumulate work $W(t,t_0) > 0$ ($<0$). $\rho(t)$ is constant only when there is no work of random forces in average.

If the average work $\overline{W(t)}$ does not depend on time over a process, we can write $W(t,t_0) = \alpha(t-t_0)$ and:

$$\rho(t) = \rho(0)\exp[-\lambda t] \qquad (20)$$

where $\lambda = \gamma\alpha$ is a constant and a Lyapunov-like exponent characterizing the variation of the distances between the state points.

## 5. Boltzmann *H* Theorem

The Boltzmann *H* function can be defined in discrete coarse graining way in phase space by:

$$H(t) = \sum_{x,P} p(x,P,t)\ln p(x,P,t) \qquad (21)$$

where the sum is over all the accessible coarse grained phase domain and $p(x,P,t) \propto \rho(x,P,t)d\Omega$ is the probability that a system is found in a phase cell of volume $d\Omega$ situated at the point $(x,P)$ at time *t*.

We have:

$$p(x,P,t) = p(x_0,P_0)\exp[-\gamma W(t,t_0)] \qquad (22)$$

It is straightforward to see that the variation of *H* function from $t_0$ to *t* is given by $\Delta H = H(t) - H(t_0) = -\gamma W(t,t_0)$. Since $\gamma$ is positive, it is necessary to prove that the cumulate work $W_R(t,t_0)$ is positive in order to yield Boltzmann *H* theorem $\Delta H \leq 0$.

Here we only provide a proof for $W(t,t_0) \geq 0$ in the case of ideal gas in which each molecule can be considered as a particle in random motion. The motion of the ensemble of particles is also random, perturbed by internal fluctuation.

Let us consider the free expansion of ideal gas between two equilibrium states at time $t_1$ and $t_2$ with volume $V_1$ and $V_2$, respectively. Let $p(x_1,P_1,t_1)$ and $p(x_2,P_2,t_2)$ be the two equilibrium distributions. We use the entropy expression:

$$S(t) = -\sum_{x,P} p(x,P,t)\ln p(x,P,t) \qquad (23)$$

For the two equilibrium states, it is straightforward to calculate (let the Boltzmann constant $k_B = 1$):

$$\Delta S = S_2 - S_1 = N\ln\frac{V_2}{V_1} = -\zeta(t_2,t_1) = \gamma W(t_2,t_1) \qquad (24)$$

Hence the variation of *H* function is $\Delta H = H_2 - H_1 = -\Delta S = -N\ln\frac{V_2}{V_1}$, meaning that $W(t_2,t_1) = \frac{N}{\gamma}\ln\frac{V_2}{V_1} \geq 0$. This is a proof for the *H* theorem in the special case of ideal gas expansion between two equilibrium states. For the motion between any two states, the property $W(t,t_0) \geq 0$ needs a more general proof from its definition above Equation (19).

## 6. Concluding Remarks

We have studied the theoretical consequences of the path probability exponentially depending on action which was an observed result in the numerical experiment of random motion of Hamiltonian systems. We have indicated that this probability was an emblem of the stochastic mechanics based on a generalization of the least action principle of the classical Hamiltonian/Lagrangian mechanics. Then we have proved a modification of Liouville's theorem within this generalized mechanics theory for the random dynamics: the distribution function of states in phase space is no more a constant in time as prescribed by Liouville's theorem before for deterministic dynamics of Hamiltonian systems. The Boltzmann *H* function can have time evolution allowing entropy increase. This increase can be proved to be a consequence of the work of random forces on the systems. For the special case of free expansion of ideal gas, this random work could be identified through the increase of entropy. More general proof of the *H* theorem from the definition of the work of random forces can be expected.

The essential idea of this work is to derive Boltzmann's *H* theorem, initially proposed to treat thermodynamic phenomena, within a mechanical theory for random motion and on the basis of a characteristic law of random dynamics: the path probability. Although this path probability has been found, by numerical simulation [8], only for a special and ideal random motion of Hamiltonian systems which statistically conserve their energy, the philosophy and the methodology should be suitable for discussing random dynamics and concomitant phenomena.

One of the aims of this work is to question the criticisms of the *H* theorem by Loschmidt, Poincaré and Zermelo [14–16] on the basis of the time reversible Hamiltonian mechanics and Liouville's theorem for regular and deterministic mechanical motion. Indeed, using Liouville's theorem to criticize *H* theorem, one should take it for granted that the state density is proportional to the probability distribution of states. Yet Liouville's theorem holds only for regular dynamics in which each trajectory can be *a priori* traced in time with certainty, each state has its unique moment of time to be visited by the system, and the frequency of visit of any phase (state) volume can be predicted exactly. Therefore, when Liouville's theorem holds, *a priori* there is no place for probability and entropy or other uncertainty in regular dynamics. On the other hand, all thermodynamic systems undergo random motion with more or less fluctuation and are intrinsically probabilistic and unpredictable. Without introducing probability and unpredictability into the basic laws of the Newtonian mechanics, any effort to relate the second law of thermodynamics to the Newtonian mechanics [17] must encounter the gap between the world of regular, deterministic, predictable and time reversible motion and the world of random, indeterministic, unpredictable and irreversible motion. As is well known Boltzmann has tried to prove the second law of thermodynamics from the *H*-theorem with the assumption "molecular chaos" (*Stosszahlansatz*) in formulating the collision term [18]. Boltzmann believes that from this assumption it is possible to break time-reversal symmetry, necessary for the entropy increase, and to solve the Loschmidt paradox [17], but Liouville's theorem still persists and forbids the *H* function to change in time.

This work is an effort to generalize the regular Newtonian mechanics into a stochastic mechanics theory through path probability which is a universal character and an emblem of all random dynamics. Within the formalism with the path probability depending on action, Liouville's theorem has to be modified to allow time evolution of the phase density as is shown in Equation (18). On the other hand, the time reversibility does not exist for Equation (5) or (10) or $\dot{P}_k = -\frac{\partial H_k}{\partial x} + R_k$ since the time symmetry along any path *k* is broken down by the random force $R_k$ whose integration in opposite direction of time should be a priori different.

It should be stressed again that the present work is just a first generalization of Hamiltonian mechanics with an ideal model of random motion: a Hamiltonian system undergoing random motion without, statistically, loss of energy through dissipation. *For the time being*, this same approach, based on the least action principle, cannot be applied to dissipative systems for which this principle is no more valid. The extension of the least action principle to dissipative system is still under progress [19].

## Acknowledgments

## Conflicts of Interest